\documentclass[preprint]{emulateapj}

\shorttitle{QUASARS ON THE MAIN-SEQUENCE PLANE}
\shortauthors{Zhang et al.}

\begin{document}

\title{Distributions of quasar hosts on the galaxy main-sequence plane}

\author{Zhoujian Zhang\altaffilmark{1, 2, 3}, Yong Shi\altaffilmark{1, 2, 4},
 George H. Rieke\altaffilmark{5}, Xiaoyang Xia\altaffilmark{6}, Yikang Wang\altaffilmark{1, 2}, Bingqing Sun\altaffilmark{7},
Linfeng Wan\altaffilmark{1,2}}

\altaffiltext{1}{School of Astronomy and Space Science, Nanjing University, Nanjing 210093, China}
\altaffiltext{2}{Key Laboratory of Modern Astronomy and Astrophysics (Nanjing University), Ministry of Education, Nanjing 210093, China}
\altaffiltext{3}{Current address: Institute for Astronomy, University of Hawaii, 2680 Woodlawn Drive, Honolulu, HI 96822, USA}
\altaffiltext{4}{Collaborative Innovation Center of Modern Astronomy and Space Exploration, Nanjing 210093, China}
\altaffiltext{5}{Department Of Astronomy And Steward Observatory, University of Arizona, 933 N Cherry Ave, Tucson, AZ 85721, USA}
\altaffiltext{6}{Tianjin Astrophysics Center, Tianjin Normal University, Tianjin 300387, China}
\altaffiltext{7}{National Astronomical Observatories, Chinese Academy of Sciences, Beijing 100012, China}

\begin{abstract}

The relation between star formation rates and stellar masses, i.e. the
galaxy main sequence, is a  useful diagnostic of galaxy evolution.  We
present  the  distributions  relative  to  the  main  sequence  of  55
optically-selected PG and 12 near-IR-selected 2MASS quasars at $z \leq
0.5$.  We  estimate the quasar  host stellar masses from  {\it Hubble}
Space Telescope or ground-based AO  photometry, and the star formation
rates   through  the   mid-infrared  aromatic   features  and   far-IR
photometry.  We find that PG quasar hosts more or less follow the main
sequence defined  by normal  star-forming galaxies while  2MASS quasar
hosts lie systematically above the main sequence. PG and 2MASS quasars
with higher  nuclear luminosities  seem to  have higher  specific SFRs
(sSFRs), although there is a large scatter. No trends are seen between
sSFRs  and SMBH  masses,  Eddington ratios  or  even morphology  types
(ellipticals, spirals and mergers).  Our results could be placed in an
evolutionary scenario with quasars emerging during the transition from
ULIRGs/mergers  to ellipticals.   However,  combined  with results  at
higher redshift, they suggest that  quasars can be widely triggered in
normal galaxies as long as they  contain abundant gas and have ongoing
star formation.

\end{abstract}                                                    
\keywords{infrared: galaxies --- galaxies: active ---  galaxies: starburst}

\section{INTRODUCTION} 

\label{sec:introduction}

Super-massive  black-holes  (SMBHs)  are  now  known  to  be  integral
components of galaxies.  The tight correlation between SMBH masses and
their  host galaxy  properties \citep[e.g.][]{Kormendy+2013},  and the
similar mass growth rates between  SMBHs and galaxies with cosmic time
\citep{Mullaney+2012},   indicate   coevolution    between   the   two
\citep{Heckman+2014}.  Quasi-stellar objects (QSOs), the manifestation
of dramatic accretion onto SMBHs, are the key stage of SMBH growth and
represent  an important  phase in  the evolution  of massive  galaxies
\citep{Sanders+1988, Hopkins+2006}.  Probing  the properties of quasar
hosts,    e.g.     host    morphology    \citep{Dunlop+2003},    color
\citep{Jahnke+2004}, interstellar  medium \citep{Xia12,  Petric15} and
star  formation  behavior   \citep{Shi+2007,  Shi+2014,  Xu+2015},  is
essential to  understanding the environment  where SMBHs grow,  and to
gain insights into the circumstances influencing coevolution.

 Color-magnitude diagrams  are frequently employed to  investigate AGN
 host properties  \citep{Silverman+2008, Xue10}.  However,  the colors
 and   magnitudes  of   galaxies  can   be  affected   by  extinction,
 metallicity, the age  of the stellar populations,  and star formation
 rates as well as  by the stellar mass. On the  other hand, the galaxy
 main sequence  characterizes the relationship between  star formation
 rates (SFRs) and stellar  masses (M$_{\star}$) of normal star-forming
 galaxies \citep{Brinchmann+2004,  Daddi07}.  Elliptical galaxies
   lie about  $\sim 1.5$ dex below  the typical SFRs of  main sequence
   galaxies  at a  given stellar  mass, while  starbursts are  located
   significantly  above ($\gtrsim  0.6$  dex) it  \citep{Whitaker12}.
 The  redshift evolution  of  the  main sequence  shows  the roles  of
 increased gas  content and  more efficient star  formation triggering
 mechanisms at high redshift  \citep{Whitaker12}.  Therefore, the main
 sequence investigation that studies the  locations of galaxies in the
 SFR  vs.   stellar  mass  plane,   is  less  degenerate  in  physical
 interpretation than color-magnitude approaches.

Previous  studies of  AGN relative  to the  main sequence  of galaxies
mainly  focus on  deep  survey data,  and thus  on  the high-z  regime
\citep{Xue10,   Mullaney+2012,   Harrison12,  Rosario+2013,   Xu+2015,
  Mullaney15},    or   on    moderate-luminosity   AGN    at   low-$z$
\citep[e.g.][]{Shimizu15}. Instead, we  will quantify the distribution
along     the      main     sequence     of      the     Palomar-Green
\citep[PG;][]{Schmidt+1983}   and   Two    Micron   All   Sky   Survey
\citep[2MASS;][]{Cutri+2001,   Smith+2002}   archetypal   low-redshift
quasar samples.   This study builds  on that of  \citet{Shi+2014}, who
presented infrared  spectroscopic and  photometric observations  of PG
and 2MASS  quasars and used them  to derive star formation  rates.  We
complement these  results by estimating  the stellar masses  using the
optical/near-IR  photometric measurements  of  quasar  hosts from  the
literature  based on  {\it  Hubble} Space  Telescope and  ground-based
adaptive optics (AO) observations.

\section{Sample}
\label{sec:sample}

\begin{figure*}[t]
\begin{center}
\includegraphics[height=2.0in]{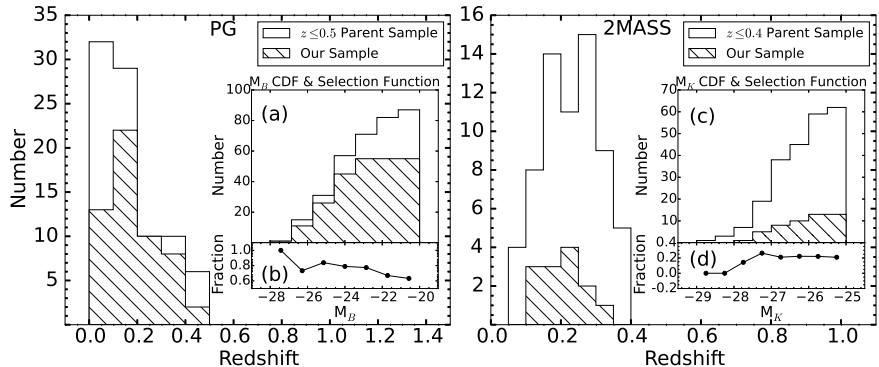}
\caption{Redshift distributions of PG (\emph{left}) and 2MASS (\emph{right})
  samples adopted in this study (\emph{diagonal hatches}), as compared with
  the parent samples (\emph{white area}). The subplots show the cumulative distribution of their absolute magnitudes (B-band for PG and K-band for 2MASS sample) and the selection function.}
\label{fig:redshift_distribution}
\end{center}
\end{figure*}

We drew sources for  our study from the parent samples  of  88 PG
quasars at $z \leq 0.5$ and  63 2MASS quasars at $z \leq 0.4$. PG
quasars are  representative of  bright optically selected  quasars and
are all classified as type 1.  2MASS quasars represent a redder
quasar  population compared  to PG  quasars but  have similar  K$_{\rm
  s}$-band luminosity \citep{Smith+2002}. The 2MASS quasars contain a
broad  range  of  types  including 21 type 1, 37 intermediate, 4 type 2
and 1 low-ionization emission line object.

To find objects  suitable for determination of host  galaxy masses, we
searched the HST  archive. Our final sample contains 56  PG objects at
$z \leq 0.5$ and 13 2MASS objects at $z \leq 0.4$.   53 PG and all
  2MASS sample have HST images, which we prefer due to the better PSF
stability;   we  also  include three  PG  objects  which  were
observed  only  by the  Gemini-N  or  the  Subaru Telescope  with  AO \citep{Guyon+2006}.
Although we  included only a small  number of 2MASS objects,  they are
representative of  the parent sample  both in redshift  and luminosity
\citep{Marble+2003, Shi+2014}.

As  shown   in  Figure  \ref{fig:redshift_distribution},   our  sample
contains $\sim 63\%$ of the PG  and  $\sim 20\%$ of the
2MASS parent sources.   The inserted plots show that our  PG and 2MASS
quasars are biased toward the  bright members of their parent samples; they are typically at $z \sim 0.16$ 
and $z \sim 0.21$ (median), respectively.
For comparison, we also selected: a complete sample of $118$ ULIRGs with 
\citep[$f_{60\mu m} > 1$ Jy, ][]{Kim+1998}  at $z \sim 0.1$, plus $587,673$ SDSS galaxies
\citep{Brinchmann+2004}.

\section{Derivation of the stellar mass and SFRs of  quasar hosts}
\label{sec:mass_sfr_host}

The stellar  masses of our quasar hosts are  calculated from host
galaxy absolute magnitudes reported in the literature. In all cases,
these are based on decompositions of 2D images into host and nuclear
components.   The       result        is       listed       in
Table 1.   Note  all literature  measurements
are rescaled to the cosmology at  $H_{0}$ = 70 km s$^{-1}$ Mpc$^{-1}$,
$\Omega_{M}$ = 0.3 and $\Omega_{\lambda}$ = 0.7.

Before deriving stellar masses,  we examine the
consistency of the measurements  from the seven studies we utilize. Overall,   their methods used 
to remove  the quasar  contamination are very similar. The image was decomposed into the quasar and 
host light with a two-dimensional  analysis, with a
stellar image or  TinyTim-produced PSF image representing the quasar contribution.  The 
host component was usually represented by a single de  Vaucouleurs (elliptical) and  a single
exponential  (disk) model, although sometimes an additional  de Vaucouleurs  plus
exponential (bulge+disk)  were added.  The final result was determined by 
$\chi^{2}$ minimization.  
For  the  PG  sample,  \citet{Veilleux+2009}  confirmed  that  their HST
measurements   were    in   good    agreement   with the prior ground-based and HST ones from
\citet{McLeod+2001},    \citet{Guyon+2006},   \citet{Hamilton+2008},
which contain $\sim$  95\% of our sample (see their  Figure 7).  The
only   exception   is   that  the   AO  measurements from \citet{Guyon+2006}
overestimate the host magnitudes by $\sim$  0.3 mag, which we corrected 
for sources  from this study. Since the 2MASS
host  galaxies are  typically brighter  than their
nuclei  by  $\sim$  0.3  mag,   it  is  reasonable  to  believe  the
reliability of the decompositions.

We  estimated  the  stellar  masses as  follows.  First, an absolute luminosity  was calculated from 
this magnitude.  For the 46 PG  quasars with $H$
  or $K$  band data, the magnitude was  converted to the luminosity  at the
  band of the magnitude with  no K-correction.  The optical magnitudes
  of the remaining 10 PG and  all 13 2MASS sources were converted into
   $r$-band  and  $i$-band  luminosities  with  appropriate  K-corrections, 
based on  the  SWIRE  template
  library \citep{Polletta+2007}. Given the lack  of the optical color, we
  derived two K-corrections for each  object corresponding to
  elliptical  and spiral SED templates.  We chose the elliptical template with an age
  of 5  Gyr and  the Sb  galaxy template  for spirals.  The different SEDs
  result  in  at most a $7\%$  difference in derived mass.   Second, stellar  masses  were
  estimated  based on  the  luminosities. \citet{Xu+2015b} demonstrate the accuracy of 
such single-band mass estimates compared with those from more complete photometry.  For H-  and
  K-band    luminosities,   we    adopted   the   logarithmic
  stellar-mass/light  ratio   of  -  0.017  and   -0.08,  respectively
  \citep{Bell+2003}.  For hosts with  optical luminosities, since we do not 
have an optical  color we  derived  two stellar  masses using  two
  mass-to-light ratios, one  appropriate for blue galaxies  and the other
  for elliptical  galaxies.  With  the  SDSS galaxies  from the 
  MPA/JHU catalog  with stellar  mass above 10$^{10}$  M$_{\odot}$, we
  found the distribution  of the mass-to-light ratios  can be described
  by  a combination  of  two  Gaussians: log($M_{\star}/M_{\odot}$)  +
  $M_{\lambda}$/2.5 = 2.23$\pm0.09$ for  red and $1.99\pm$0.18 blue in
  $r$-band, and  $2.1\pm0.07$ for  red and  $1.89\pm0.15$ for  blue in
  $i$-band, where $M_{\lambda}$ is the absolute magnitude in AB.

 The final quoted errors are based on uncertainties in the host galaxy
 photometry  and in  the stellar  M/L.  For  the first,  we adopt  the
 results  from the  works reporting  the host  photometry, since  they
 performed  detailed error  analysis taking  account of  the residuals
 from PSF subtraction. For the M/L, we adopt the typical uncertainties
 in  the NIR  ($\sim 0.2$  dex), and  optical bands  ($\sim 0.08$  for
 ellipticals, and  $\sim 0.17$  dex for spirals  from the  SDSS galaxy
 distributions) to describe the scatter of the conversion. For nine PG
 quasars   with   dynamical   measurements   of   host   galaxies   by
 \citet{Dasyra07}, we found that our photometric masses are lower than
 the dynamical masses by only a  factor of 1.2 (median), indicating no
 large systematic offset in our mass estimates.

The SFRs were obtained  from \cite{Shi+2014}.  Basically, the aromatic
features or  160 $\mu$m  are used  to estimate the  SFRs based  on the
star-forming templates  provided by  \cite{Rieke+2009}.   In cases
  where  we have  70um photometry  and at  least one  of (i)  aromatic
  features or  (ii) 160um photometry,  we use the latter  to calibrate
  the relationship between  70um photometry and SFR. We  then use this
  relationship  to calculate  SFRs in  cases where  we only  have 70um
  photometry.  The SFR estimates of the  sample in this paper are the
following: (1) $11.3$ $\mu$m aromatic features (1 object); (2) MIPS 70
$\mu$m  photometry (18  objects);  and  (3) MIPS  or  PACS 160  $\mu$m
photometry  (48  objects).  However,  the  weak  host galaxy  infrared
emission from  PG2349-014 and 2MASS  J023430.6+243835 made it  hard to
estimate their SFRs, so we eliminated them from subsequent analysis.

\section{Results}
\label{sec:results}

\subsection{Distribution of quasars along the main sequence}

\begin{figure*}[t]
\begin{center}
  \includegraphics[height=3.0in]{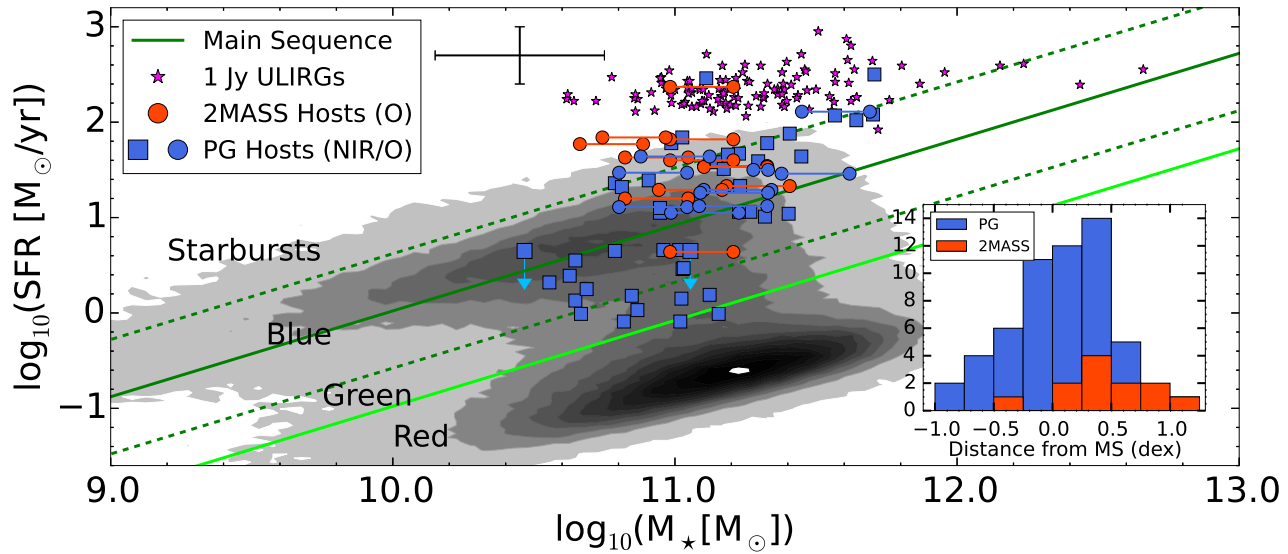}
  \caption{SFR--M$_{\star}$  relation  for  67 2MASS (\emph{red})
  and  PG  (\emph{blue})   quasar hosts.  Galaxies 
      with NIR measurements appear as squares, while those with optical 
      photometry are shown as two circles corresponding to two K-corrections 
      (see text). The typical error-bar is also noted.
    Over-plotted data include $118$ complete $1$ Jy ULIRGs (\emph{pink
      stars})  and  $587,673$  SDSS DR7  galaxies  (\emph{filled  grey
      contours}). Contours  in grey scale  show the number  density of
    the  SDSS  sample, and  demonstrate  a  bimodal distribution.  The
    \emph{dark  green   solid  line}   describes  the   main  sequence
    \citep{Peng+2010} with  two dashed lines  0.6 dex above  and below
    the sequence; the \emph{lime solid line} is visually determined as
    the boundary between green valley  and red galaxies, that is about
    10  times  below  the  sequence.  In  addition,  two  \emph{ \emph{light blue}
      downward arrows} are for PG  0026+129 and PG1121+422, whose SFRs
    are $3\sigma$ values. The subplot  at the right lower corner shows
    the distributions of the distance of our 2MASS (red) and PG (blue)
    objects to the main sequence. }
\label{fig:DR7_Jy_Sample}
\end{center}
\end{figure*}

The   distribution   of   our   67  quasar  host galaxies   in SFR vs. stellar mass is   shown   in   Figure
\ref{fig:DR7_Jy_Sample} with the  $1$ Jy ULIRGs and  SDSS DR7 galaxies
(see Section \ref{sec:sample}) overlaid. The stellar masses of the $1$
Jy  ULIRGs are  from their  K$'$ band  magnitudes \citep{Kim+2002} as 
discussed  in Section  \ref{sec:mass_sfr_host}, and
their   SFRs  are   estimated  from   the  far-infrared   luminosities
\citep{Kim+1998}     based    on     the    scaling     relation    by
\cite{Kennicutt+1998}.  The  SFRs  and  stellar  masses  of  the  SDSS
galaxies are from  the SDSS MPA/JHU catalog, multiplied by 1.5 to convert to 
the Salpeter  IMF.  We used the  quantitative fitting
of SDSS galaxies from \cite{Peng+2010} adjusted to the Salpeter
IMF to define  the main sequence. The upward and  downward boundaries are 
 set to  be $\pm$ 0.6 dex,  which is  twice the  1-$\sigma$
scatter  of  the main  sequence  \citep{Daddi07}.   The dividing  line
between the ``green valley'' and  red galaxies is estimated by eye,
roughly 1.0 dex  below the main sequence.

As  shown in  Figure~\ref{fig:DR7_Jy_Sample}, the majority of  PG quasar  hosts reside within the
main sequence, with 23\% in the starburst regime and 11\% in the green
valley. The  median distance  of these galaxies to  the main  sequence is
small, only  0.07 dex. The  standard deviation of these  distances is
about 0.41  dex. Although this  is larger than the  1-$\sigma$ scatter
(0.3 dex) of the main sequence itself,  the large measured errors of the specific
SFRs of  the PG quasar hosts  could broaden the  intrinsic scatter.  PG quasar host galaxies
thus seem to follow more or less the main sequence.

2MASS  quasar  host galaxies,  on  the  other hand,  systematically  lie  above  the
main sequence, with only  1 out of  12 objects below.  The
median distance of 2MASS quasar hosts to  the main sequence is  0.43 dex with a
standard deviation of   0.33 dex. Super-giants in starbursts can  contribute 
to  the  NIR bands  \citep{Shier96},  possibly making  the NIR-selected  2MASS
sources biased toward ones with higher SFRs.

\subsection{sSFR of quasar hosts as a function of global SMBH and host properties}

\begin{figure*}
\begin{center}
\includegraphics[height=3.0in]{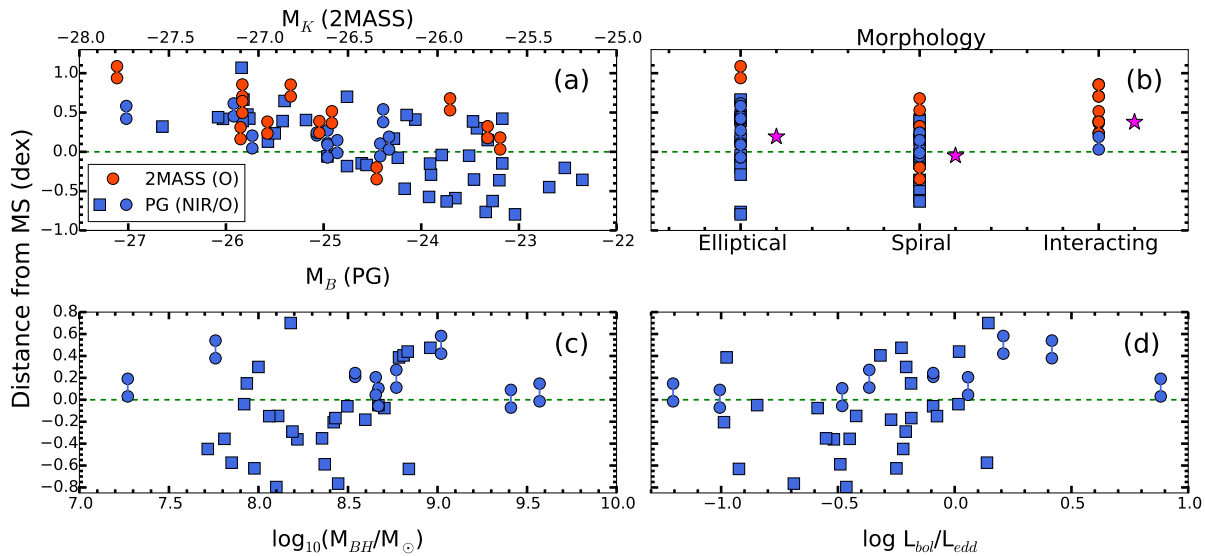}
\caption{ Distance to the MS as a function of (a) nuclear luminosities for PG (with NIR/optical photometry) at $B$-band and 2MASS  (optical) at $K$-band, (b) morphology types, (c)  SMBH masses for PG quasars  and  (d) Eddington ratios for PG quasars.}
\label{fig:sSFR_global}
\end{center}
\end{figure*}

We here investigate  the specific SFR (sSFR  $=$ SFR$/$M$_{\star}$) of
quasar hosts  as functions of  global SMBH and host  properties.  SMBH
masses  are  extracted  from  \cite{Veilleux+2009}  by  averaging  the
results from different methods, including  spheroid luminosity,
    spheroid  velocity dispersion,  reverberation mapping,  and virial
    relation.   A  bolometric  correction from  \cite{Elvis+1994}  was
    assumed     for     Eddington     ratios.      As     shown     in
    Figure~\ref{fig:sSFR_global}  (a),  despite   the  large  scatter,
    higher nuclear luminosity  quasars seem to have  on average higher
    sSFRs, e.g., at $M_{\rm B}$ $<$ -25 the majority of PG quasars lie
    above  the sequence  with  the  median 0.42  dex  higher than  the
    sequence, while  lower $B$-band  luminosity PG  quasars distribute
    around the  main sequence with a  median value 0.11 dex  below the
    sequence.  A recent work by \citet{Shimizu15} points out that 
      the  distribution of  moderate-luminosity AGN  at low-$z$  
      peaks $\sim  0.6$ dex below  the main sequence.  It  seems that
    their result represents an extension of our finding of higher SFRs
    in      higher       nuclear      luminosity       AGN      hosts.
    Figure~\ref{fig:sSFR_global}  (b)   shows  no   relationship  with
    morphologies (elliptical, spirals and merging).  sSFRs are neither
    apparently enhanced in major  mergers nor suppressed in elliptical
    hosts. No  relationships are  seen with  SMBH masses  or Eddington
    ratio   as   shown   in   Figure~\ref{fig:sSFR_global}   (c)   and
    Figure~\ref{fig:sSFR_global} (d), respectively.

\subsection{Redshift evolution of the sSFR of quasar hosts}\label{sec:sSFR_z}

Figure~\ref{fig:sSFR_z} shows  the redshift evolution of  the sSFRs of
our quasar hosts along with  those of IR-selected high-z quasars from
\citet{Xu+2015}, providing further evidence that  the PG quasar hosts follow more or less
the evolution  of the main  sequence. At $z$ = 0.3-0.5, PG  quasar hosts have
some bias toward  higher sSFRs,  in contrast to those  below $z$ = 0.3.   At similar
redshifts   (0.3-0.5),  quasars   from   \citet{Xu+2015}  have   lower
luminosities  than ours  but show  lower  SFRs. This suggests that  it is  the
luminosity   instead of  the  redshift  that  causes  PG
quasar hosts to tend to have higher sSFRs at $z$ = 0.3-0.5.

\section{Discussion: comparisons with other studies}
\label{sec:discussion}

\begin{figure*}
\begin{center}
\includegraphics[height=2.8in]{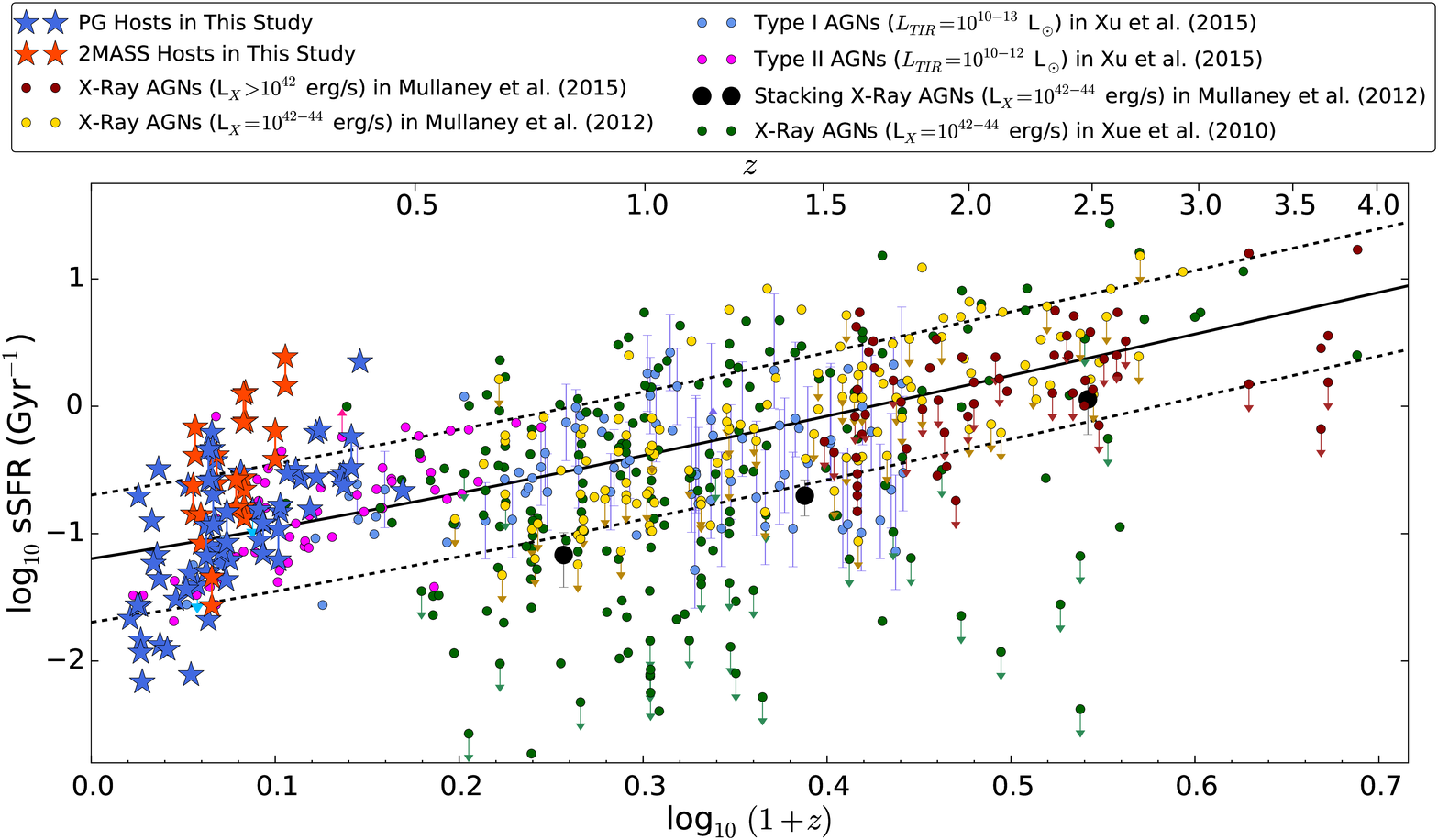}
\caption{Redshift evolution of sSFRs of our quasar hosts along with measurements from the literature. }
\label{fig:sSFR_z}
\end{center}
\end{figure*}

\citet{Xu+2015} investigated  SFRs and
stellar masses of  24$\mu$m-selected quasars up to z = 1.8. For about  half of their sample,
SFRs can  be measured based  on individual detections in  the far-IR.
As with our study, their  sample is  composed  of high-luminosity
AGN, i.e.  quasars, and has enough objects ($\sim$  300) to
reach statistically robust conclusions.  They found that the
quasar hosts roughly follow the main sequence to z = 1.8.

\citet{Xue10} compared color and SFRs of moderate luminosity X-ray, narrow-line AGN
hosts to galaxies of similar stellar mass but without AGN, and found  at $z$ $>$1 that the two stellar-matched-mass 
samples have similar SFRs. However, below $z$=1  they found 2-3  times  higher SFRs  
for the galaxies with AGN. Nonetheless, as shown in Figure~\ref{fig:sSFR_z}, the host galaxies in
their sample tend to fall on or below the main sequence, with a minority in the starburst regime, including those at z $<$ 1.
 \citet{Mullaney+2012} used {\it Herschel}  100 and 160
$\mu$m  measurements to  investigate  the SFRs of  moderate
luminosity X-ray AGN ($L_{\rm x}$=10$^{42-44}$  erg/s) at $z$ = 0.5-3. 
Their color selection to identify cases where the stellar output is dominant (allowing accurate  
host mass estimates) strongly favors type 2 AGN, so their final sample resembles that of \citet{Xue10}.
Based  on  the individually detected  far-IR  sources  combined  with  a  stacking
analysis, they found  that the SFRs  of these AGN hosts roughly follow the  main sequence with
slightly   lower    (20\%)   average SFR  values.  This  combined  with their  findings  of  no
relationship  between  SFRs and  SMBH  accretion  rates support  their
suggestion that moderately  luminous AGN are unlikely  to be triggered
by major mergers  instead of internal processes, and that  there may be no
causal   link  between   nuclear   activity   and  star   formation. 
\citet{Harrison12}    investigated    X-ray-luminous   AGN    ($L_{\rm
  x}$ $>$ 10$^{44}$ erg/s) and found their average SFRs are not suppressed
or enhanced compared to star-forming  galaxies at the same redshift. A
similar  result is  found  for  optically-selected and  X-ray-selected
quasars  between  $z$=0.5  and  $z$=2, again  based  on  a  stacking
analysis  \citep{Rosario+2013}.  By   decomposing  the  SED  including
upper-limits  to   measured  SFRs  for  individual   X-ray  sources  at
$z$=0.2-2.5,    \citet{Stanley15}     reach    similar    conclusions.

 The above studies based on  deep surveys consistently found that the more luminous hosts
of AGN follow more  or  less  the  main sequence  of  star-forming  galaxies.  However, for purely star-forming 
galaxies of similar mass ($>10^{10} M_\odot$), the MS is a rough upper limit for the SFR and 
many galaxies fall well below it \citep{Brinchmann+2004}. Similarly, AGN studies 
that measure the SFR individually in a substantial fraction of the observed sample find that the stacked 
results for the remainder also fall substantially below the MS \citep[e.g.,][]{Xu+2015b}.  Studies that 
rely more heavily on stacking may miss this behavior.  In confirmation, \citet{Mullaney15} 
observed some of their  X-ray AGN hosts with ALMA and found a median SFR
that is 0.4 dex below the main sequence. This result is similar to the low-redshift
moderate-luminosity AGN that systematically lie below the sequence \citep{Shimizu15}.
The overall result is that AGNs of moderate and high luminosity reside in  host galaxies that
are not dramatically different from  star-forming galaxies  in general,  and major
mergers may not be the main mechanism to trigger star formation in AGN
hosts.


We can  also place our finding in the widely adopted picture of major-mergers
driving both star formation and AGN  \citep{Sanders+1988,   Hopkins+2006},   as  it   is
consistent  with  the  evolutionary  sequence from  ULIRGs  to  type-2
quasars to type-1  quasars to elliptical in which  sSFRs decrease from
ULIRGs to obscured  quasars (2MASS) to unobscured quasars  (PG) to red
dead ellipticals.  However,  the fact that PG quasars lie  on the main
sequence could  also indicate that  the triggering of quasars  can happen
widely  in all  types  of  galaxies with  cold  gas  and ongoing  star
formation. Even in major mergers, the phase with enhanced SFRs is only
a fraction of  the whole duration of the interaction \citep{DiMatteo05},  and quasars may
be triggered in all phases of  major mergers, which is consistent with
no observed relationship between sSFRs and morphology types as seen in
Figure~\ref{fig:sSFR_global} (b).

\section{Conclusions}\label{sec:summary}

In summary, from  the distribution of quasar hosts  on the main-sequence plane, we found that  PG quasars follow more  or less
the  main sequence.   Our  small  sample of  2MASS  quasar hosts  lies
systematically above  the sequence,  a result  possibly affected  by a
selection  bias.  The  behavior of  the  PG quasars  at z  $< 0.5$  is
similar to that for samples studied by  others at z $> 0.5$.  While no
apparent  relationships  are  observed between  sSFRs  and  black-hole
masses, Eddington ratio or even  morphology types, quasars with higher
nuclear luminosities  show on  average higher  sSFRs despite  a larger
scatter. Although our finding is consistent with the merger-driven AGN
scenario  that links  ULIRGs to  type-2 quasars  to type-1  quasars to
ellipticals, it may also imply that quasars can be widely triggered in
host  galaxies with  rich gas  and ongoing  star formation,  without a
close connection to major mergers.

\acknowledgements

We thank the anonymous referee  for constructive comments that improve
the paper significantly. Z.Z.  and  Y.S.  acknowledge support for this
work  from the  National Natural  Science Foundation  of China  (grant
11373021), the  Strategic Priority  Research Program The  Emergence of
Cosmological  Structures of  the  Chinese Academy  of Sciences  (grant
No. XDB09000000), and Excellent Youth Foundation of Jiangsu Scientific
Committee (grant  BK20150014).  Z.Z.  also  thanks for the  support by
the NSFC grant J1210039.


\clearpage

\clearpage
\LongTables
\begin{deluxetable*}{lllllllllllllllllllll}
\tablewidth{0pc}
\tabletypesize{\tiny}
\tablecaption{Photometry, stellar masses and SFRs of PG \& 2MASS Sample}
\tablehead{
\colhead{}		&	\colhead{}		&	\colhead{}		&	\colhead{}		&	\colhead{}		&	\colhead{}			&	\colhead{}			&	\colhead{}					&	\colhead{host}		&	\colhead{}	&	\colhead{$R_{e}$}		&	\colhead{}		&	\colhead{log M$_{\star}$}		&	\colhead{log SFR}	\\
\colhead{Name}		&	\colhead{$z$}			&	\colhead{Morp.}		&	\colhead{Inst.}			&	\colhead{Band}			&	\colhead{$M_{\rm host}$}	&	\colhead{$M_{\rm nuc}$}	&	\colhead{$m_{\rm total}$}	&	\colhead{fraction}	&	\colhead{$\sigma_{\rm host}$}		&	\colhead{(kpc)}					&	\colhead{Ref.}		&	\colhead{($M_{\odot}$)}		&	\colhead{($M_{\odot}$ yr$^{-1}$)}			\\
\colhead{(1)}			&	\colhead{(2)}			&	\colhead{(3)}			&	\colhead{(4)}			&	\colhead{(5)}			&	\colhead{(6)}			&	\colhead{(7)}			&	\colhead{(8)}			&	\colhead{(9)}				&	\colhead{(10)}				&	\colhead{(11)}		&	\colhead{(12)}		&	\colhead{(13)}			&	\colhead{(14)}		}
\startdata
PG 0007+106		&	0.089	&	E		&	NICMOS	&	H	&	-24.26	&	-24.58	&	...		&	...		&	0.07	&	2.97		&	5		&	11.02		&	0.66		\\
PG 0026+129		&	0.142	&	E		&	NICMOS	&	H	&	-24.36	&	-25.82	&	...		&	...		&	0.22	&	0.90		&	5		&	11.06		&	$<$0.65	\\		
PG 0043+039		&	0.385	&	E		&	WFPC2	&	V	&	-21.94	&	-24.76	&	...		&	...		&	0.20	&	...		&	2, 4		&	[10.88, 11.12]	&	1.64		\\	
PG 0050+124		&	0.061 	&	B+D		&	NICMOS	&	H	&	-24.96	&	-25.08	&	...		&	...		&	0.01	&	1.03		&	5		&	11.30		&	1.59		\\
PG 0052+251		&	0.155 	&	S		&	WFPC2	&	V	&	-22.35	&	-23.07	&	...		&	...		&	0.20	&	...		&	2, 4		&	[10.99, 11.23]	&	1.05		\\
PG 0157+001		&	0.163	&	S		&	WFPC2	&	V	&	-23.61	&	-22.57	&	...		&	...		&	0.20	&	...		&	2, 4		&	11.71		&	2.50		\\
				&			&	E		&	NICMOS	&	H	&	-25.99	&	-25.76	&	...		&	...		&	0.20	&	1.82		&	5		&				&			\\
PG 0804+761		&	0.100	&	Dp		&	$\star$	&	H	&	-23.77	&	-26.09	&	...		&	...		&	0.26	&	...		&	7		&	10.82		&	-0.09		\\
PG 0838+770		&	0.131	&	B+D		&	NICMOS	&	H	&	-25.02	&	-23.95	&	...		&	...		&	0.03	&	0.56		&	5		&	11.32		&	1.01		\\
PG 0844+349		&	0.064 	&	B+D		&	NICMOS	&	H	&	-23.89	&	-24.32	&	...		&	...		&	0.02	&	0.28		&	5		&	10.87		&	0.03		\\
PG 0923+201		&	0.190 	&	E		&	WFPC2	&	V	&	-22.27	&	-24.00	&	...		&	...		&	0.20	&			&	2, 4		&	11.27		&	1.06		\\
				&			&	E		&	NICMOS	&	H	&	-24.89	&	-26.20	&	...		&	...		&	0.14	&	1.30		&	5		&				&			\\
PG 0947+396		&	0.206 	&	D		&	NICMOS	&	H	&	-23.99	&	-25.69	&	...		&	...		&	0.12	&	...		&	1		&	10.91		&	1.39		\\
PG 0953+414		&	0.234 	&	S		&	WFPC2	&	V	&	-22.49	&	-24.50	&	...		&	...		&	0.20	&	...		&	2, 4		&	[11.10, 11.34]	&	1.29		\\
PG 1001+054		&	0.161	&	E		&	NICMOS	&	H	&	-23.64	&	-25.27	&	...		&	...		&	0.18	&	2.38		&	5		&	10.65		&	0.55		\\
PG 1004+130		&	0.240	&	S		&	WFPC2	&	V	&	-23.33	&	-24.89	&	...		&	...		&	0.20	&	...		&	2, 4		&	[11.38, 11.62]	&	1.46		\\
PG 1012+008		&	0.185 	&	S, I		&	WFPC2	&	V	&	-22.46	&	-23.02	&	...		&	...		&	0.20	&	...		&	2, 4		&	[11.09, 11.33]	&	1.26		\\
PG 1048+342		&	0.167 	&	E		&	NICMOS	&	H	&	-24.09	&	-24.29	&	...		&	...		&	0.22	&	...		&	1		&	10.95		&	1.05		\\	 
PG 1100+772		&	0.315	&	...		&	NICMOS	&	J	&	...		&	...		&	14.45	&	0.270	&	0.22	&	...		&	6		&	11.45		&	1.64		\\
				&			&	...		&	NICMOS	&	H	&	...		&	...		&	13.75	&	0.223	&	0.22	&	...		&	6		&				&			\\
				&			&	...		&	NICMOS	&	K	&	...		&	...		&	12.96	&	0.180	&	0.22	&	...		&	6		&				&			\\
PG 1116+215		&	0.177	&	S		&	WFPC2	&	V	&	-22.69	&	-24.46	&	...		&	...		&	0.20	&	...		&	2, 4		&	10.79		&	0.65		\\
				&	0.176	&	?		&	NICMOS	&	H	&	-23.99	&	-27.31	&	...		&	...		&	0.51	&	...		&	5		&				&			\\
PG 1119+120		&	0.050 	&	B+D		&	NICMOS	&	H	&	-23.84	&	-23.49	&	...		&	...		&	0.02	&	0.45		&	5		&	10.85		&	0.18		\\
PG 1121+422		&	0.224 	&	E		&	NICMOS	&	H	&	-22.89	&	-25.69	&	...		&	...		&	0.22	&	...		&	1		&	10.47		&	<0.65	\\
PG 1126--041		&	0.060	&	B+D		&	NICMOS	&	H	&	-24.30	&	-24.94	&	...		&	...		&	0.09	&	...		&	5		&	11.03		&	0.47		\\
PG 1151+117		&	0.176 	&	D		&	NICMOS	&	H	&	-23.29	&	-25.09	&	...		&	...		&	0.12	&	...		&	1		&	10.63		&	0.39		\\
PG 1202+281		&	0.165 	&	E		&	WFPC2	&	V	&	-21.89	&	-22.30	&	...		&	...		&	0.20	&	...		&	2, 4		&	[10.80, 11.04]	&	1.11		\\
PG 1211+143		&	0.081	&	...		&	$\star$	&	H	&	-23.11	&	-25.48	&	...		&	...		&	0.28	&	...		&	7		&	10.56		&	0.32		\\
PG 1216+069		&	0.331	&	E		&	WFPC2	&	V	&	-21.82	&	-25.03	&	...		&	...		&	0.20	&	...		&	2, 4		&	11.03		&	1.84		\\
				&			&	...		&	NICMOS	&	J	&	...		&	...		&	14.51	&	0.100	&	0.22	&	...		&	6		&				&			\\
				&			&	...		&	NICMOS	&	H	&	...		&	...		&	13.92	&	0.090	&	0.22	&	...		&	6		&				&			\\
				&			&	...		&	NICMOS	&	K	&	...		&	...		&	13.43	&	0.100	&	0.22	&	...		&	6		&				&			\\
PG 1226+023		&	0.158	&	E		&	WFPC2	&	V	&	-23.51	&	-26.46	&	...		&	...		&	0.20	&	...		&	2, 4		&	[11.45, 11.69]	&	2.11		\\
PG 1229+204		&	0.064 	&	S		&	WFPC2	&	V	&	-21.60	&	-21.54	&	...		&	...		&	0.20	&	...		&	2, 4		&	11.12		&	0.19		\\	
				&			&	B+D		&	NICMOS	&	H	&	-24.53	&	-23.71	&	...		&	...		&	0.10	&	3.65		&	5		&				&			\\
PG 1259+593		&	0.477	&	...		&	NICMOS	&	J	&	...		&	...		&	14.77	&	0.080	&	0.22	&	...		&	6		&	11.17		&	1.51		\\
				&			&	...		&	NICMOS	&	H	&	...		&	...		&	13.88	&	0.064	&	0.22	&	...		&	6		&				&			\\
				&			&	...		&	NICMOS	&	K	&	...		&	...		&	13.04	&	0.050	&	0.22	&	...		&	6		&				&			\\
PG 1302--102		&	0.278	&	?		&	NICMOS	&	H	&	-25.24	&	-26.81	&	...		&	...		&	0.01	&	...		&	5		&	11.41		&	1.88		\\
PG 1307+085		&	0.155 	&	E		&	WFPC2	&	V	&	-21.69	&	-23.60	&	...		&	...		&	0.20	&	...		&	2, 4		&	10.96		&	0.66		\\		
				&			&	E		&	NICMOS	&	H	&	-24.12	&	-25.37	&	...		&	...		&	0.17	&	1.32		&	5		&				&			\\
PG 1309+355		&	0.184	&	S		&	WFPC2	&	V	&	-22.89	&	-23.80	&	...		&	...		&	0.20	&	...		&	2, 4		&	11.40		&	1.04		\\
				&			&	E		&	NICMOS	&	H	&	-25.53	&	-25.89	&	...		&	...		&	0.06	&	3.49		&	5		&				&			\\
PG 1322+659		&	0.168 	&	E		&	NICMOS	&	H	&	-23.69	&	-25.29	&	...		&	...		&	0.22	&	...		&	1		&	10.79		&	1.36		\\
PG 1351+640		&	0.088	&	?		&	$\star$	&	K$'$	&	-23.94	&	-25.99	&	...		&	...		&	0.28	&	...		&	7		&	10.81		&	1.32		\\
PG 1352+183		&	0.158 	&	D		&	NICMOS	&	H	&	-23.39	&	-24.59	&	...		&	...		&	0.12	&	...		&	1		&	10.67		&	-0.01		\\
PG 1354+213		&	0.300 	&	E		&	NICMOS	&	H	&	-24.59	&	-25.29	&	...		&	...		&	0.22	&	...		&	1		&	11.15		&	1.58		\\
PG 1402+261		&	0.164 	&	...		&	WFPC2	&	V	&	-21.89	&	-23.39	&	...		&	...		&	0.20	&	...		&	2, 4		&	[10.80, 11.04]	&	1.47		\\
PG 1411+442		&	0.090 	&	?		&	NICMOS	&	H	&	-24.28	&	-25.30	&	...		&	...		&	0.07	&	...		&	5		&	11.02		&	0.15		\\
PG 1425+267		&	0.366	&	E		&	WFPC2	&	V	&	-22.72	&	-24.76	&	...		&	...		&	0.20	&	...		&	2, 4		&	11.33		&	1.78		\\
				&			&	...		&	NICMOS	&	J	&	...		&	...		&	15.11	&	0.280	&	0.22	&	...		&	6		&				&			\\
				&			&	...		&	NICMOS	&	H	&	...		&	...		&	14.30	&	0.214	&	0.22	&	...		&	6		&				&			\\
				&			&	...		&	NICMOS	&	K	&	...		&	...		&	13.54	&	0.180	&	0.22	&	...		&	6		&				&			\\
PG 1426+015		&	0.086 	&	S		&	WFPC2	&	V	&	-22.44	&	-23.14	&	...		&	...		&	0.20	&	...		&	2, 4		&	11.23		&	1.06		\\
				&			&	B+D		&	NICMOS	&	H	&	-24.79	&	-24.98	&	...		&	...		&	0.06	&	1.64		&	5		&				&			\\
PG 1427+480		&	0.221 	&	D		&	NICMOS	&	H	&	-24.09	&	-25.19	&	...		&	...		&	0.22	&	...		&	1		&	10.95		&	1.10		\\	
PG 1435--067		&	0.126	&	E		&	NICMOS	&	H	&	-23.44	&	-24.74	&	...		&	...		&	0.02	&	...		&	5		&	10.69		&	0.25		\\	
PG 1440+356		&	0.079 	&	B+D		&	NICMOS	&	H	&	-24.80	&	-25.27	&	...		&	...		&	0.01	&	0.39		&	5		&	11.23		&	1.33		\\
PG 1444+407		&	0.267 	&	S		&	WFPC2	&	V	&	-23.08	&	-24.41	&	...		&	...		&	0.20	&	...		&	2, 4		&	[11.28, 11.33]	&	1.50		\\
PG 1512+370		&	0.371 	&	E		&	WFPC2	&	V	&	-22.93	&	-24.65	&	...		&	...		&	0.20	&	...		&	2, 4		&	11.19		&	1.66		\\
				&			&	...		&	NICMOS	&	J	&	...		&	...		&	15.48	&	0.290	&	0.22	&	...		&	6		&				&			\\
				&			&	...		&	NICMOS	&	H	&	...		&	...		&	14.68	&	0.224	&	0.22	&	...		&	6		&				&			\\
				&			&	...		&	NICMOS	&	K	&	...		&	...		&	13.57	&	0.130	&	0.22	&	...		&	6		&				&			\\
PG 1543+489		&	0.400	&	...		&	NICMOS	&	J	&	...		&	...		&	15.12	&	0.150	&	0.22	&	...		&	6		&	11.11		&	2.46		\\
				&			&	...		&	NICMOS	&	H	&	...		&	...		&	14.17	&	0.105	&	0.22	&	...		&	6		&				&			\\
				&			&	...		&	NICMOS	&	K	&	...		&	...		&	13.11	&	0.060	&	0.22	&	...		&	6		&				&			\\
PG 1545+210		&	0.264 	&	E		&	WFPC2	&	V	&	-22.60	&	-24.21	&	...		&	...		&	0.20	&	...		&	2, 4		&	[11.09, 11.33]	&	1.12		\\
PG 1613+658		&	0.129 	&	?		&	NICMOS	&	H	&	-25.83	&	-25.91	&	...		&	...		&	0.14	&	...		&	5		&	11.64		&	2.02		\\
PG 1617+175		&	0.112 	&	?		&	NICMOS	&	H	&	-23.34	&	-25.40	&	...		&	...		&	0.01	&	...		&	5		&	10.65		&	0.13		\\
PG 1626+554		&	0.133 	&	E		&	NICMOS	&	H	&	-24.27	&	-25.29	&	...		&	...		&	0.20	&	5.53		&	5		&	11.02		&	-0.09		\\
PG 1700+518		&	0.292 	&	?		&	NICMOS	&	H	&	-25.64	&	-27.85	&	...		&	...		&	0.08	&	...		&	5 		&	11.57		&	2.07		\\
PG 1704+608		&	0.372 	&	S		&	WFPC2	&	V	&	-23.86	&	-25.43	&	...		&	...		&	0.20	&	...		&	2, 4		&	11.70		&	2.08		\\
				&			&	...		&	NICMOS	&	J	&	...		&	...		&	14.08	&	0.250	&	0.22	&	...		&	6		&				&			\\	
				&			&	...		&	NICMOS	&	H	&	...		&	...		&	13.40	&	0.217	&	0.22	&	...		&	6		&				&			\\	
				&			&	...		&	NICMOS	&	K	&	...		&	...		&	12.39	&	0.140	&	0.22	&	...		&	6		&				&			\\	
PG 2130+099		&	0.063 	&	B+D		&	NICMOS	&	H	&	-24.29	&	-25.02	&	...		&	...		&	0.23	&	2.83		&	5		&	11.03		&	0.46		\\
PG 2214+139		&	0.066 	&	E		&	NICMOS	&	H	&	-24.61	&	-24.49	&	...		&	...		&	0.14	&	2.78		&	5		&	11.16		&	-0.01		\\
PG 2233+134		&	0.325 	&	D		&	NICMOS	&	H	&	-24.19	&	-25.89	&	...		&	...		&	0.22	&	...		&	1		&	10.99		&	1.78		\\
PG 2251+113		&	0.326 	&	?		&	NICMOS	&	H	&	-24.79	&	-27.74	&	...		&	...		&	0.01	&	...		&	5		&	11.23		&	1.67		\\	
PG 2349--014		&	0.174	&	?		&	NICMOS	&	H	&	-25.88	&	-26.35	&	...		&	...		&	0.05	&	...		&	5		&	11.66		&	...		\\
2M J005055.7+293328		&	0.136	&	S 	&	WFPC2	&	I	&	-22.45	&	-20.45	&	...	&	...	&	1.00	&	...		&	3		&	[10.82, 11.05]	&	1.20		\\	
2M J015721.0+171248		&	0.213 	&	? 	&	WFPC2	&	I	&	-22.05	&	-20.85	&	...	&	...	&	1.00	&	...		&	3		&	[10.66, 10.89]	&	1.77		\\
2M J023430.6+243835		&	0.310 	&	... 	&	WFPC2	&	I	&	-23.25	&	-24.05	&	...	&	...	&	1.00	&	...		&	3		&	[11.14, 11.37]	&	...		\\
2M J034857.6+125547		&	0.210 	&	? 	&	WFPC2	&	I	&	-22.25	&	-19.55	&	...	&	...	&	1.00	&	...		&	3		&	[10.74, 10.97]	&	1.84		\\
2M J125807.4+232921		&	0.259 	&	? 	&	WFPC2	&	I	&	-22.45	&	-23.75	&	...	&	...	&	1.00	&	...		&	3		&	[10,82, 11.05]	&	1.63		\\
2M J130700.6+233805		&	0.275	&	E 	&	WFPC2	&	I	&	-22.85	&	-18.95	&	...	&	...	&	1.00	&	...		&	3		&	[10.98, 11.21]	&	2.37		\\
2M J145331.5+135358		&	0.139	&	S 	&	WFPC2	&	I	&	-22.85	&	-19.65	&	...	&	...	&	1.00	&	...		&	3		&	[10.98, 11.21]	&	1.82		\\
2M J163700.2+222114		&	0.211	&	S 	&	WFPC2	&	I	&	-23.15	&	-21.65	&	...	&	...	&	1.00	&	...		&	3		&	[11.10, 11.33]	&	1.54		\\
2M J165939.7+183436		&	0.170	&	?	&	WFPC2	&	I	&	-22.85	&	-22.45	&	...	&	...	&	1.00	&	...		&	3		&	[10.98, 11.21]	&	1.60		\\
2M J171442.7+260248		&	0.163	&	S 	&	WFPC2	&	I	&	-22.85	&	-22.45	&	...	&	...	&	1.00	&	...		&	3		&	[10.98, 11.21]	&	0.64		\\
2M J222221.1+195947		&	0.211	&	? 	&	WFPC2	&	I	&	-22.75	&	-23.35	&	...	&	...	&	1.00	&	...		&	3		&	[10.94, 11.17]	&	1.29		\\		
2M J222554.2+195837		&	0.147	&	S 	&	WFPC2	&	I	&	-23.35	&	-17.95	&	...	&	...	&	1.00	&	...		&	3		&	[11.18, 11.41]	&	1.33		\\
2M J234449.5+122143		&	0.199	&	... 	&	WFPC2	&	I	&	-23.15	&	-23.45	&	...	&	...	&	1.00	&	...		&	3		&	[11.10, 11.33]	&	1.53		
\enddata
\tablecomments{Column  (1):  the  name  of the  quasars;  Column  (2):
  redshift; Column (3): morphology (E =  Elliptical, S = Spiral, B+D =
  Bulge+Disk,  Dp =  Disk present,  ?  = Ambiguous);  Column (4):  the
  instrument used  for HST  observations,  and
  $\star$  for ground-based  AO observations;  Column (5):  magnitudes
  bands; Column  (6): the  absolute magnitudes  of the  host galaxies;
  Column (7): the absolute magnitude  of the nuclei. Note both Column
  (6)  and   (7)  have  been   rescaled  into  the   same  cosmology,
  i.e. H$_{0}$  = 70 km s$^{-1}$  Mpc $^{-1}$, $\Omega_{M} =  0.3$ and
  $\Omega_{\Lambda}  = 0.7$;  Column (8):  total apparent  magnitudes;
  Column (9): the  host fraction; Column (11): photometric uncertainty of 
  host magnitudes; Column (11):  the effective radius;
  Column (12): references,  1.McLeod \& McLeod (2001);  2. Hamilton et
  al.  (2002); 3.  Marble  et al.  (2003);
  4. Hamilton  et al. (2008); 5.  Veilleux et al. (2009);  6. Shang et
  al.  (2011);  7.  Guyon  et al.  (2006); Column (13): stellar masses calculated in this paper, the ones with two values are derived from optical photometry with two mass-to-light ratios (see text); Column (14): SFR from Shi et al. (2014).}   
\label{tab:photometry_sample}
\end{deluxetable*}

\end{document}